%% file: main.tex
\documentclass[sigplan,screen]{acmart}

\setcopyright{acmcopyright}
\acmPrice{15.00}
\acmYear{2022}
\copyrightyear{2022}
\acmSubmissionID{pldiws22mapsmain-p12-p}
\acmISBN{978-1-4503-9273-0/22/06}
\acmConference[MAPS '22]{Proceedings of the 6th ACM SIGPLAN International Symposium on Machine Programming}{June 13, 2022}{San Diego, CA, USA}
\acmBooktitle{Proceedings of the 6th ACM SIGPLAN International Symposium on Machine Programming (MAPS '22), June 13, 2022, San Diego, CA, USA}

\usepackage{microtype}
\usepackage{algorithmic}
\usepackage{caption}
\usepackage{stfloats} 
\usepackage{multirow}
\usepackage{fancybox}
\usepackage{graphicx}
\usepackage{lipsum}
\usepackage{xspace}
\usepackage{xcolor}
\usepackage{comment}
\usepackage{balance}
\usepackage{enumitem}
\usepackage{hyperref}
\usepackage{cleveref}

\usepackage[labelfont={bf}, font={small}]{caption}
\usepackage{subcaption}
\usepackage{listings}
\usepackage{color}

\definecolor{pblue}{rgb}{0.13,0.13,1}
\definecolor{pgreen}{rgb}{0,0.5,0}
\definecolor{pred}{rgb}{0.9,0,0}
\definecolor{pgrey}{rgb}{0.46,0.45,0.48}

\title{Syntax-Guided Program Reduction for Understanding Neural Code Intelligence Models}

\author{Md Rafiqul Islam Rabin}
\email{mrabin@uh.edu}
\affiliation{%
  \institution{University of Houston}
  \city{Houston}
  \state{TX}
  \country{USA}
}

\author{Aftab Hussain}
\email{ahussain27@uh.edu}
\affiliation{%
  \institution{University of Houston}
  \city{Houston}
  \state{TX}
  \country{USA}
}

\author{Mohammad Amin Alipour}
\email{maalipou@central.uh.edu}
\affiliation{%
  \institution{University of Houston}
  \city{Houston}
  \state{TX}
  \country{USA}
}



\input{mystyle}

\begin{document}

\input{abstract}
\maketitle

\input{introduction}
\input{related}
\input{methodology}
\input{result}
\input{validity}
\input{conclusion}
\input{impact}

\balance
\bibliography{refs}
\bibliographystyle{ACM-Reference-Format}

\end{document}

%% file: mystyle.tex
\newcommand{\Part}[1]{\noindent\textbf{#1}}

\cornersize{.2}
\newcounter{observation}
\newcommand{\observation}[1]{\refstepcounter{observation}
        \begin{center}
        \vspace{2pt}
        \Ovalbox{
            \begin{minipage}{0.9\columnwidth}
                \textbf{Observation \arabic{observation}:} #1
            \end{minipage}
        }
        \vspace{2pt}
        \end{center}
}

\newcommand{\ctv}{\textsc{Code2Vec}\xspace}
\newcommand{\cts}{\textsc{Code2Seq}\xspace}

\newcommand{\JL}{\textsc{Java-Large}\xspace}
\newcommand{\JTT}{\textsc{Java-Top10}\xspace}

\newcommand{\mnp}{\textsc{MethodName}\xspace}

\newcommand{\ddd}{Delta Debugging\xspace}
\newcommand{\sivand}{\textsc{Sivand}\xspace}
\newcommand{\sivanddd}{\textsc{Sivand-DD}\xspace}
\newcommand{\sivandperses}{\textsc{Sivand-Perses}\xspace}

\newcommand{\typeF}{\textsc{Frequent}\xspace}
\newcommand{\typeR}{\textsc{Rare}\xspace}
\newcommand{\typeS}{\textsc{Small}\xspace}
\newcommand{\typeL}{\textsc{Large}\xspace}

\newcommand{\ci}{code intelligence\xspace}


%% file: abstract.tex
\begin{abstract}
Neural code intelligence (CI) models are opaque black-boxes and offer little insight on the features they use in making predictions. This opacity may lead to distrust in their prediction and hamper their wider adoption in safety-critical applications. Recently, input program reduction techniques have been proposed to identify key features in the input programs to improve the transparency of CI models. However, this approach is syntax-unaware and does not consider the grammar of the programming language. 

In this paper, we apply a syntax-guided program reduction technique that considers the grammar of the input programs during reduction. Our experiments on multiple models across different types of input programs show that the syntax-guided program reduction technique is faster and provides smaller sets of key tokens in reduced programs. We also show that the key tokens could be used in generating adversarial examples for up to 65\% of the input programs.
\end{abstract}

%% file: introduction.tex
\section{Introduction}
\label{sec:introduction}


A neural code intelligence (CI) model is a deep neural network that takes a program as input and predicts certain properties of that program as output, e.g., predicting method name \cite{allamanis2015suggesting}, variable name \cite{allamanis2018ggnn}, or type \cite{hellendoorn2018type} from a program body. Unfortunately, these models are opaque black-boxes and it is hard to interpret and debug the results of these models.
Recently, there is an increasing interest in probing CI models to improve their transparency and identify and understand their potential flaws. For instance, recent studies suggest that state-of-the-art CI models do not always generalize to other experiments \cite{kang2019generalizability, rabin2020evaluation}; they heavily rely on specific individual tokens \cite{compton2020obfuscation, rabin2021sivand} or structures \cite{rabin2020demystifying, rabin2021code2snapshot}, are prone to learn noise \cite{rabin2021memorization} or duplication \cite{allamanis2019duplication}, and are often vulnerable to semantic-preserving adversarial examples \cite{rabin2019tnpa, yefet2020adversarial, rabin2021generalizability}.

Neural CI models, as in other neural models, represent an input program as continuous distributed vectors that are computed after training on large volumes of programs. It makes understanding what input features a model has learned a very challenging task. For example, \ctv model \cite{alon2019code2vec} learns to represent an input program as a single fixed-length high dimensional embedding, however, it is difficult to understand the meaning or characteristics of each of these dimensions.
Attention-based approaches are commonly applied to find important code elements in a program. For example, \citet{bui2019autofocus} attempt to identify relevant code elements by perturbing statements of the program and combining corresponding attention and confidence scores. However, the attention-based approach poorly correlates with key code elements.

Several studies have already been conducted to find relevant input features in models' inferences. \citet{allamanis2015suggesting} use a set of features from programs and show that extracting relevant features that capture global context is essential for learning effective code context. \citet{rabin2020demystifying} attempt to find key input features of a label by manually inspecting some input programs of that label. However, this manual inspection does not scale to large datasets with many labels. 
\citet{suneja2021probing} and \citet{rabin2021sivand} use input program reduction techniques, based on \ddd \cite{zeller2002dd}, to find minimal inputs that preserve the model's prediction, hence finding key tokens in the program with respect to the prediction.
However, these approaches are syntax-unaware and can create a large number of invalid programs as they do not follow the grammar of the programming language during the reduction, which in turn can hinder the identification of the key tokens.

In this paper, we apply a syntax-guided reduction technique, called \sivandperses, to remove irrelevant parts from an input program.
Given a CI model and an input program, our approach adopts Perses \cite{sun2018perses} to reduce the input program while preserving the model's prediction. The approach continues reducing the input program as long as the model maintains the same prediction on the reduced program as on the original program.
As the syntax-guided technique follows the syntax of the programming language, it will always generate valid input programs. Using the syntax information also can improve the performance of the reduction, as the reduction only explores the space of syntactically valid programs.
The main insight is that, by reducing some input programs of a target label, we may better understand what key input features a CI model learns from the training dataset.

An experiment with two CI models and four types of input programs suggests that the syntax-guided \sivanddd outperforms the syntax-unaware alternative \sivanddd proposed in ~\cite{rabin2021sivand}. While \sivandperses always generates valid programs, \sivanddd generates only around $10\%$ valid programs. On average, \sivandperses removes $20\%$ more tokens, takes $70\%$ fewer reduction steps, and spends half of the reduction time compared to \sivanddd for reducing an input program.
Furthermore, our results show that we can find key input features by reducing input programs using \sivandperses, which can provide additional explanation for a prediction and highlight the importance of input features in reduced programs by triggering $10\%$ more mispredictions with $50\%$ fewer adversarial examples.

\smallskip
\noindent\textbf{Contributions.}
This paper makes the following contributions:
\begin{itemize}
    \item We propose a syntax-guided program reduction approach, \sivandperses, for the identification of key features in neural code intelligence models.
    \item We evaluate the performance of \sivandperses and compare it with the state-of-the-art technique on two CI models for the code summarization task.
\end{itemize}

%% file: related.tex
\section{Related Work}
\label{sec:related}

There has been some work in the area of code intelligence that focuses on the understanding of what relevant features a black-box model learns for correct predictions.
While some works \cite{rabin2019tnpa, compton2020obfuscation, rabin2020evaluation, kang2019generalizability, yefet2020adversarial, rabin2021generalizability, suneja2021probing, rabin2021sivand} study the reliance of models on specific features, many works \cite{allamanis2015suggesting, bui2019autofocus, rabin2020demystifying, suneja2021probing, rabin2021sivand, wang2021demystifying} focus on finding relevant features for explaining models' prediction.

\subsection{Learning Representation of Source Code}
An input program is usually represented as vector embeddings for processing and analyzing by neural models. \citet{allamanis2014learning} introduced a framework that processed token sequences and abstract syntax trees of code to represent the raw programs. \citet{alon2019code2vec} proposed an attention-based neural model that uses a bag of path-context from abstract syntax tree for representing any arbitrary code snippets. \citet{allamanis2018ggnn} constructed data and control flow graphs from programs to encode a code snippet. \citet{hellendoorn2018type} proposed an RNN-based model using sequence-to-sequence type annotations for type suggestion.
There are some surveys on the taxonomy of models that exploit source code analysis \cite{allamanis2018survey,sharma2021survey}. \citet{chen2019embeddings} also provide a survey that includes the usage of code embeddings based on different granularities of programs. However, these models are often black-box and do not provide any insight on the meaning or characteristic of learned embeddings. What features or patterns these embeddings represent are largely unknown. In this work, we extract key input features that a model learns for predicting a target label as an explanation of learned embeddings.

\subsection{Reliance on Specific Features}
Models often learn irrelevant features, simple shortcuts, or even noise for achieving target performance.
\citet{compton2020obfuscation} show that the code2vec embeddings highly rely on variable names and cannot embed an entire class rather than an individual method. They investigate the effect of obfuscation on improving code2vec embeddings that better preserves code semantics. They retrain the code2vec model with obfuscated variables to forcing it on the structure of code rather than variable names and aggregate the embeddings of all methods from a class.
Following the generalizability of word embeddings, \citet{kang2019generalizability} assess the generalizability of code embeddings in various software engineering tasks and demonstrate that the learned embeddings by code2vec do not always generalize to other tasks beyond the example task it has been trained for.
\citet{rabin2021generalizability} and \citet{yefet2020adversarial} demonstrate that the models of code often suffer from a lack of robustness and are vulnerable to adversarial examples. They mainly introduce small perturbations in code for generating adversarial examples that do not change any semantics and find that the simple renaming, adding or removing tokens changes model's predictions.
\citet{suneja2021probing} uncover the model’s reliance on incorrect signals by checking whether the vulnerability in the original code is missing in the reduced minimal snippet. They find that model captures noises instead of actual signals from the dataset for achieving high predictions. \citet{rabin2021sivand} demonstrates that models often use just a few simple syntactic shortcuts for making prediction. \citet{rabin2021memorization} later show that models can fit noisy training data with excessive parameter capacity and thus suffer in generalization performance.
As models often learn noise or irrelevant features for achieving high prediction performance, the lack of understanding of what input features models learn would hinder the trustworthiness to correct classification. Such opacity is substantially more problematic in critical applications such as vulnerability detection or automated defect repair. In this work, we extract key input features for CI models in order to provide better transparency and explanation of predictions.

\subsection{Extracting Relevant Input Features}
Several kinds of research have been done in finding relevant input features for models of source code.
\citet{allamanis2015suggesting} exhibit that extracting relevant features is essential for learning effective code context. They use a set of hard-coded features from source code that integrate non-local information beyond local information and train a neural probabilistic language model for automatically suggesting names. However, extracting hard-coded features from source code may not be available for arbitrary code snippets and in dynamically typed languages.
\citet{bui2019autofocus} propose a code perturbation approach for interpreting attention-based models of source code. It measures the importance of a statement in code by deleting it from the original code and analyzing the effect on predicted outputs. However, the attention-based approach often poorly correlates with key elements and suffers from a lack of explainability.
\citet{rabin2020demystifying} attempt to find key input features of a label by manually inspecting some input programs of that label. They extract handcrafted features for each label and train simple binary SVM classification models that achieve highly comparable results to the higher dimensional code2vec embeddings for the method naming task. However, the manual inspection cannot be applied to a large dataset.
\citet{wang2021demystifying} recently propose a mutate-reduce approach to find key features in the code summarization models considering valid programs.
\citet{suneja2021probing} and \citet{rabin2021sivand} apply a syntax-unaware program reduction technique, \ddd \cite{zeller2002dd}, to find the minimal snippet which a model needs to maintain its prediction. By removing irrelevant parts to a prediction from the input programs, the authors aim to better understand key features in the model inference. 
However, the syntax-unaware approach creates a large number of invalid programs during the reduction as it does not follow the syntax of programs, thus increases the total steps and time of reduction. In this work, we apply a syntax-guided program reduction technique that overcomes the overhead raised by the syntax-unaware technique.

%% file: methodology.tex
\section{Design and Implementation}
\label{sec:design}

\input{resources/workflow}

This section describes our approach of extracting input features for code intelligence (CI) models by syntax-guided program reduction. We use Perses \cite{sun2018perses} as the syntax-guided program reduction technique in our study. We first provide an overview of how Perses works and then describe how we adopt it in the workflow of our \sivandperses approach.

\medskip
\Part{\textsc{Perses}.}
\citet{sun2018perses} have proposed the framework for syntax-guided program reduction called Perses. 
Given an input program, the grammar of that programming language, and the output criteria, Perses reduces the input program with respect to the grammar while preserving the output criteria. It mainly follows the steps below:
\begin{itemize}
    \item It first parses the input program into a parse tree by normalizing the definition of grammar. 
    \item Then it traverses the tree and determines whether a tree node is deletable (e.g. whether it follows the grammar and preserves the output criteria). If yes, it prunes the sub-tree from that node and generates a valid reduced program, else it ignores that node and avoids generating invalid programs. Thus, in each iteration of reduction, it ensures generating syntactically valid program variants that preserves the same output criteria. 
    \item Next, the deletion of one node may enable the deletion of another node. Therefore, Perses is repeatedly applied to the reduced program until no more tree nodes can be removed -- a process known as fixpoint mode reduction.
    \item The final reduced program is called 1-tree-minimal, and any further attempts to reduce the program would generate an invalid program or change the output criteria.
\end{itemize}

For supporting a programming language data, the syntax-guided technique leverages knowledge about program syntax for avoiding generating syntactically invalid programs. We integrate Perses as a black-box framework into \sivandperses for extracting input features of CI models.

\medskip

\Part{Workflow.}
Figure~\ref{fig:approach} depicts a high-level view of the workflow in \sivandperses.
Given a set of input programs, our approach reduces each input program using Perses while preserving the same prediction by the CI model. 
The approach removes irrelevant parts from an input program and keeps the minimal code snippet that the CI model needs to preserve its prediction. 
The main insight is that, by reducing some input programs of a target label, we can identify key input features of the CI model for that target label.
Our approach follows the steps below:
\begin{itemize}
    \item Given an input program $P$ and a CI model $M$, our approach first records the prediction $y$ (i.e., predicted method name) given by the CI model $M$ on the input program $P$, such as $y = M(P)$.
    \item Using Perses, we then generate a candidate reduced program $R'$ by removing some nodes from the tree of the input program $P$, such as $R' = Perses(P)$.
    \item If the candidate reduced program $R'$ does not hold the same prediction $y$ by the CI model $M$ (i.e., $y \neq M(R')$), we reject this candidate program and create another candidate program by removing some other nodes from the tree of the input program.
    \item If the candidate reduced program $R'$ preserves the same prediction $y$ by the CI model $M$ (i.e., $y = M(R')$), we continue reduction and iteratively search for the final reduced program $R$ that produces the same prediction $y$.
    \item The final reduced program is 1-tree-minimal, which contains the key input features that the CI model must need for making the correct prediction $y$.
\end{itemize}

After reducing a set of input programs of a target label, we extract the node type and token value from the abstract syntax tree (AST) of each reduced program.
Every extracted element from reduced programs is considered as a \emph{candidate} input feature.
The most common elements are identified as label-specific \emph{key} features and other uncommon elements are identified as input-specific \emph{sparse} features.

\medskip
\Part{Implementation.}
Our approach is model-agnostic and can be applied for various tasks and programming datasets.
In this paper, for experimentation of our approach, we study two well-known code intelligence models (\ctv and \cts), a popular code intelligence task (\mnp) and one commonly used programming language dataset (\JL) with different types of input programs. 

\paragraph{Task}
We use the method name prediction (\mnp \cite{allamanis2015suggesting}) task in this study. 
This task is commonly used by researchers in the code intelligence domain for various applications such as code summarization \cite{allamanis2015suggesting,allamanis2016summarization}, representation learning \cite{alon2019code2vec,alon2019code2seq}, neural testing \cite{kang2019generalizability,yefet2020adversarial,rabin2021generalizability}, feature extraction \cite{rabin2021sivand,wang2021demystifying}, and so on \cite{allamanis2018survey,sharma2021survey}.
In the \mnp task, a model attempts to predict the name of a method from its body. For example, given the code snippet ``\texttt{void f(int a, int b) \{int temp = a; a = b; b = temp;\}}'', the \ctv model correctly predicts the method's name as ``\texttt{swap}''.

\paragraph{Models}
We use the \ctv~\cite{alon2019code2vec} and \cts~\cite{alon2019code2seq} \ci models for \mnp task. Both models use paths from abstract syntax trees (AST) to encode a program. Given a sample expression ``\texttt{a = b;}'', an example of path context in AST is ``\texttt{a, <NameExpr ↑ AssignExpr ↓ IntegerLiteralExpr>, b}''.
\begin{itemize}
    \item \ctv. This model extracts a bag of path-contexts from the AST of the program where each path-context includes a pair of terminal nodes and the corresponding path between them. The model learns embeddings of these path-contexts during training and uses an attention mechanism to aggregate multiple path-contexts to a single code vector. The code vector is used as a representation of the program for making a prediction.
    \item \cts. This model also extracts a bag of path-contexts from the AST of the program but it sub-tokenizes each path-context. The \cts model uses a bi-directional LSTM to encode paths node-by-node, and another LSTM to decode a target sequence one-by-one.
\end{itemize}

\paragraph{Dataset}
For the \mnp task, we use the \JL dataset \cite{alon2019code2seq}. This dataset contains a total of $9,500$ Java projects from GitHub, where $9,000$ projects are for the training set, $200$ projects for the validation set, and $300$ projects for the test set. Using training set and validation set, we train both the \ctv and \cts models.

\paragraph{Input Types}
The dataset from GitHub is often imbalanced and contains different sizes and frequencies of input programs. Therefore, we choose different types of input programs from the \JL test set to evaluate the effectiveness of our approach in terms of reduction and feature extraction.

\begin{itemize}
    \item Frequent Methods: We randomly sample a total of $100$ input programs from the top-10 most occurring method names.
    \item Rare Methods: We randomly sample a total of $100$ input programs from the least occurring method names.
    \item Smaller Methods: We randomly sample a total of $100$ input programs that contains less than $10$ lines of code.
    \item Larger Methods: We randomly sample a total of $50$ input programs that have around $100$ lines of code.
\end{itemize}

Moreover, to demonstrate the key input features, we select correctly predicted input programs from the ten most frequent labels of the \JL test set for feature extraction. Those labels (methods) are: \texttt{equals}, {\tt main}, {\tt setUp}, {\tt onCreate}, {\tt toString}, {\tt run}, \texttt{hashCode}, \texttt{init}, \texttt{execute}, and \texttt{get}.

\paragraph{Syntax-unaware Reduction Technique}
We use \sivanddd \cite{rabin2021sivand} in this study which uses the \ddd algorithm as the syntax-unaware program reduction technique. \citet{zeller2002dd} have proposed the \ddd algorithm to reduce the size of an input program. The algorithm iteratively splits an input program into multiple candidate programs by removing parts of the input program. The algorithm then checks if any resulting candidate program preserves the prediction of the model on the original input program. When the algorithm finds a candidate satisfying the property, it uses the candidate as the new base to be reduced further. Otherwise, the algorithm increases the granularity for splitting, until it determines that the input program cannot be reduced further. Considering the input reduction type (token or char), we use the following two variations of \sivanddd:

\begin{itemize}
    \item \sivand-Token: In this token level reduction approach, \sivanddd reduces the size of an input program token by token. We mostly use the \sivand-Token as the default baseline for \sivanddd in this study.
    
    \item \sivand-Char: In this char level reduction approach, \sivanddd reduces the size of an input program char by char. We use the \sivand-Char approach to provide an additional explanation in \Cref{sec:results_explanation} and \Cref{code:explanation_main}.
\end{itemize}

\citet{rabin2021sivand, rabin2021artifact_dd} described in more detail how the \sivanddd technique is applied in the workflow of reducing input programs for CI models.

\medskip
\Part{Artifacts}. We have publicly shared the artifacts of this study at \href{https://github.com/mdrafiqulrabin/CI-DD-Perses}{https://github.com/mdrafiqulrabin/CI-DD-Perses}.

%% file: resources/workflow.tex
\begin{figure*}
    \centering
    \captionsetup{font=large}
    \includegraphics[width=0.9\textwidth]{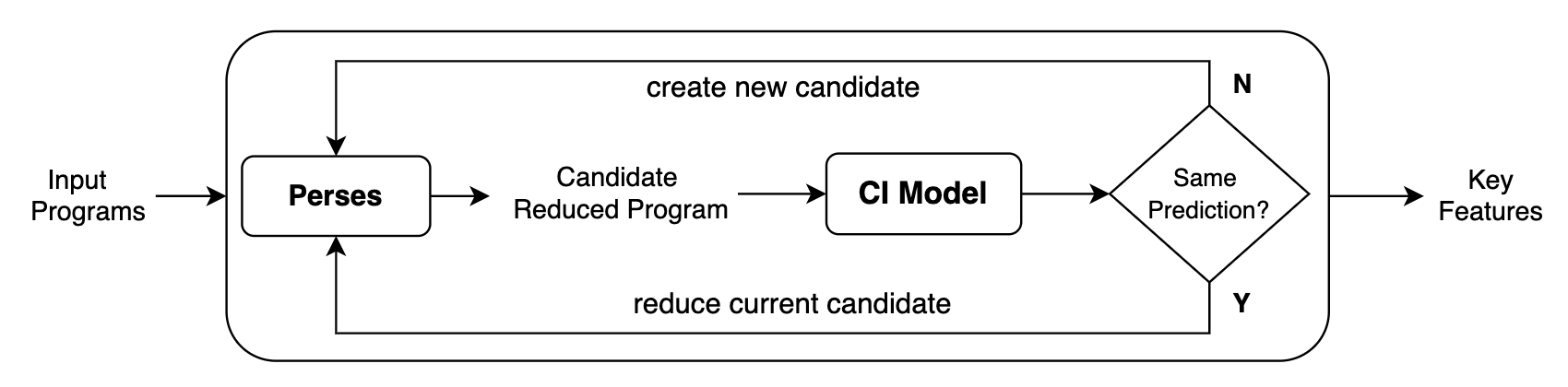}
    \caption{Workflow of \sivandperses approach.}
    \label{fig:approach}
\end{figure*}

%% file: result.tex
\section{Results}
\label{sec:results}

In this section, we present the result of our experiments on the \ctv and \cts models and the \JL dataset for different input types.

\input{results/comparison}
\subsection{Comparative Analysis}
Here, we provide a systematic comparison between the syntax-guided program reduction technique and the syntax-unaware program reduction technique.
In particular, we compare the syntax-guided \sivandperses and the syntax-unaware \sivanddd in terms of token reduction, reduction steps and reduction time.

\subsubsection{Token Reduction}
\label{sec:results_reduction}
The goal of \sivandperses and \sivanddd is to remove irrelevant tokens from an input program as much as possible while preserving the same prediction of the CI model. 
\Cref{fig:reduction} shows their such ability in reducing the size of the original input programs for different input types. We can see that, for all input types, \sivandperses reduces more tokens from input programs than \sivanddd. On average, \sivandperses removes $20\%$ more tokens from input programs than \sivanddd. The difference is most significant (around $30\%$) in \typeL input types and less significant (around $5\%$) in \typeR input types.
This result suggests that \sivandperses is more powerful than \sivanddd in reducing the size of an input program.


\subsubsection{Reduction Steps}
\label{sec:results_steps}
The reduction is applied repeatedly to an input program until finding the final minimal program, from where no more tokens can be removed. 
From \Cref{fig:steps}, we can see that \sivandperses on average can reach the final minimal program within $5$ reduction steps. However, \sivanddd makes around $20$ reductions in \typeF-\typeR-\typeS input types and more than $50$ reductions in \typeL input type, to reach the final minimal program. The \sivanddd reduces an input program by a sequence of tokens and backtracks if the reduced program is invalid, where \sivandperses can prune an entire sub-tree from the AST of the program and always generates a valid reduced program following grammar. Thus, \sivandperses takes a much lower number of reduction steps than \sivanddd to reach the final minimal program.

\subsubsection{Reduction Time}
\label{sec:results_time}
We now compare the average time taken by \sivandperses and \sivanddd for reducing an input program. As \sivanddd takes excessive invalid steps, \sivandperses is expected to spend less time for program reduction.
\Cref{fig:time} shows \sivandperses reduces an input program faster than \sivanddd, for all input types and, especially so for \typeL input type. In \typeF-\typeR-\typeS input types, both \sivandperses and \sivanddd spend less than $2$ minutes to reduce an input program and comparatively \sivandperses takes $30$ seconds less time than \sivanddd. In \typeL input types, \sivanddd spends around $17$ minutes for the reduction of a large program but \sivandperses takes only $8$ minutes, which is around $50\%$ less than \sivanddd.

\observation{\sivandperses allows more token removal than \sivanddd and always creates valid candidate programs. Compared to \sivanddd, \sivandperses is more likely to reach the final minimal program in a smaller number of reduction steps, which decreases the total reduction time.}

\input{results/feature_summary}

\input{results/feature_list}
\subsection{Key Input Features}
\label{sec:results_features}

Here, we provide the summary of extracted input features that CI models learn from training dataset for predicting the target method name. In our experiment, we consider all tokens in reduced programs as \textit{candidate} tokens for input features. We define a label-specific \textit{key} feature as a candidate token that appears in at least $50\%$ of the reduced programs, where other infrequent tokens are input-specific \textit{sparse} features. 
For brevity and page limit, we only show the Top-$10$ most frequent methods in \Cref{table:feature_summary} and \Cref{table:feature_list}.

From \Cref{table:feature_summary}, considering both \ctv and \cts models, we can see that both \sivandperses and \sivanddd identify around $50$ tokens, in total, as label-specific key features in Top-$10$ methods.
However, \sivanddd contains a total of $324$ candidate tokens in reduced programs, which is $1.36$x times higher than \sivandperses that contains a total of $238$ candidate tokens.
In some methods, i.e., `equals' and `setUp', the total number of candidate tokens in \sivanddd reduced programs is almost $2$x times higher than the candidate tokens in \sivandperses reduced programs. This may suggest that the tokens found from the reduced programs of \sivanddd are more input-specific while the tokens found from the reduced programs of \sivandperses are more label-specific.

Furthermore, \Cref{table:feature_list} shows the label-specific key features (sorted by their frequency of occurrences in reduced programs) extracted by \sivanddd and \sivandperses from their reduced programs. These key input features from reduced programs can help to understand the prediction of the CI model for a target label. For example, \sivanddd and \sivandperses reveal that ``\texttt{void, args, String, Exception}'' are key features for the `main' method. It may highlight that a sample input program containing those tokens is more likely to be predicted as the `main' method by CI models. However, the key feature of the \cts model for the `init' method is only ``\texttt{void}'' as it is the only common token in the reduced programs. Moreover, we did not find any common token in the reduced programs of the `get' method for the \cts model. Thus, by extracting key features from reduced programs of a target label, we may also get an intuition about whether a CI model is learning a label correctly or not.


\observation{According to our results, \sivandperses reveals more label-specific key features in its syntax-guided reduced programs, while \sivanddd contains more input-specific sparse features in its syntax-unaware reduced programs.}

\input{results/ex_explanation_main}
\subsection{Multiple Explanations for a Specific Prediction}
\label{sec:results_explanation}

Different program simplification approaches, i.e., \sivanddd and \sivandperses, provide a different set of key features for a target label by a CI model (\Cref{table:feature_list}). Those different features can help us find multiple explanations for a specific prediction. 
For instance, \cts predicts the code snippet in \Cref{code:explanation_main}a as the `main' method. 
\sivanddd with char-based program reduction (\sivand-Char) gives the minimal program in \Cref{code:explanation_main}b, that \cts can predict as main. We can see the presence of the \texttt{Main} identifier in the method body of \Cref{code:explanation_main}b which is one of the key tokens for the target prediction. On the other hand, the \sivanddd with token-based program reduction (\sivand-Token) gives the minimal program in \Cref{code:explanation_main}c, which suggests the argument \texttt{args} has an important role in the target prediction. However, with the AST-based program reduction (\sivandperses), the minimal program is shown in \Cref{code:explanation_main}d which highlights the signature of the method, for which \cts still can predict the same target label. Having these multiple explanations can improve the transparency of the models' inferences.

\observation{Different program reduction techniques may provide additional explanations for a better understanding of a specific prediction.}

\input{results/key_adversarial}
\subsection{Key Targeted Adversarial Attacks on Models}
\label{sec:results_adverserial}

Here, we evaluate the importance of key input features in programs by evaluating the adversarial robustness \cite{yefet2020adversarial, rabin2021generalizability} of CI models where we change the extracted key features. We generate adversarial examples by applying semantic-preserving variable renaming transformation on programs, similar to \cite{rabin2021generalizability}, where we separately change each variable and all of its occurrences in the program with token \texttt{var}. We particularly compare the prediction of CI models before and after the variable renaming. In this experiment, we generate three types of adversarial sets: original adversarial set, key adversarial set, and reduced adversarial set. In original adversarial set, we target the original input programs and generate candidate transformed programs by considering all variables. In key adversarial set, we also target the original input programs but generate candidate transformed programs by considering variables that occur in the key feature list. In reduced adversarial set, we target the reduced input programs for generating candidate transformed programs. 
The results of change in prediction (misprediction) for variable renaming transformation are shown in \Cref{table:key_adversarial}. 

According to \Cref{table:key_adversarial}, on average, the number of generated candidate transformed programs in original adversarial set is around $3$x times higher than the original programs, however, only around $12\%$ of them trigger mispredictions. Next, the number of generated candidate transformed programs in key adversarial set is around $1.5$x times higher than the original programs and trigger around $22\%$ mispredictions. Although the key adversarial set contains almost $50\%$ less candidate transformed programs than the original adversarial set, they trigger around $10\%$ more mispredictions. On the other hand, the reduced programs are the minimal programs that CI models keep for preserving their target predictions. Therefore, the number of generated candidates transformed programs in reduced adversarial set is the lowest as there are fewer tokens to apply transformations. However, the transformation on reduced programs is more powerful and triggers the highest percentage of mispredictions, up to $65\%$. Moreover, comparing between \sivanddd and \sivandperses, in most cases, \sivandperses generated candidates transformed programs show a higher rate of misprediction than \sivanddd. 

\observation{The adversarial programs based on key input features trigger $10\%$ more mispredictions with $50\%$ fewer candidates. The \sivandperses generated candidate programs are more vulnerable to adversarial transformation than \sivanddd, thus, highlighting the importance of key input features in programs.}

%% file: results/comparison.tex
\begin{figure*}
\centering
\captionsetup{font=large}
\begin{subfigure}{.33\textwidth}
  \centering
  \includegraphics[width=\linewidth]{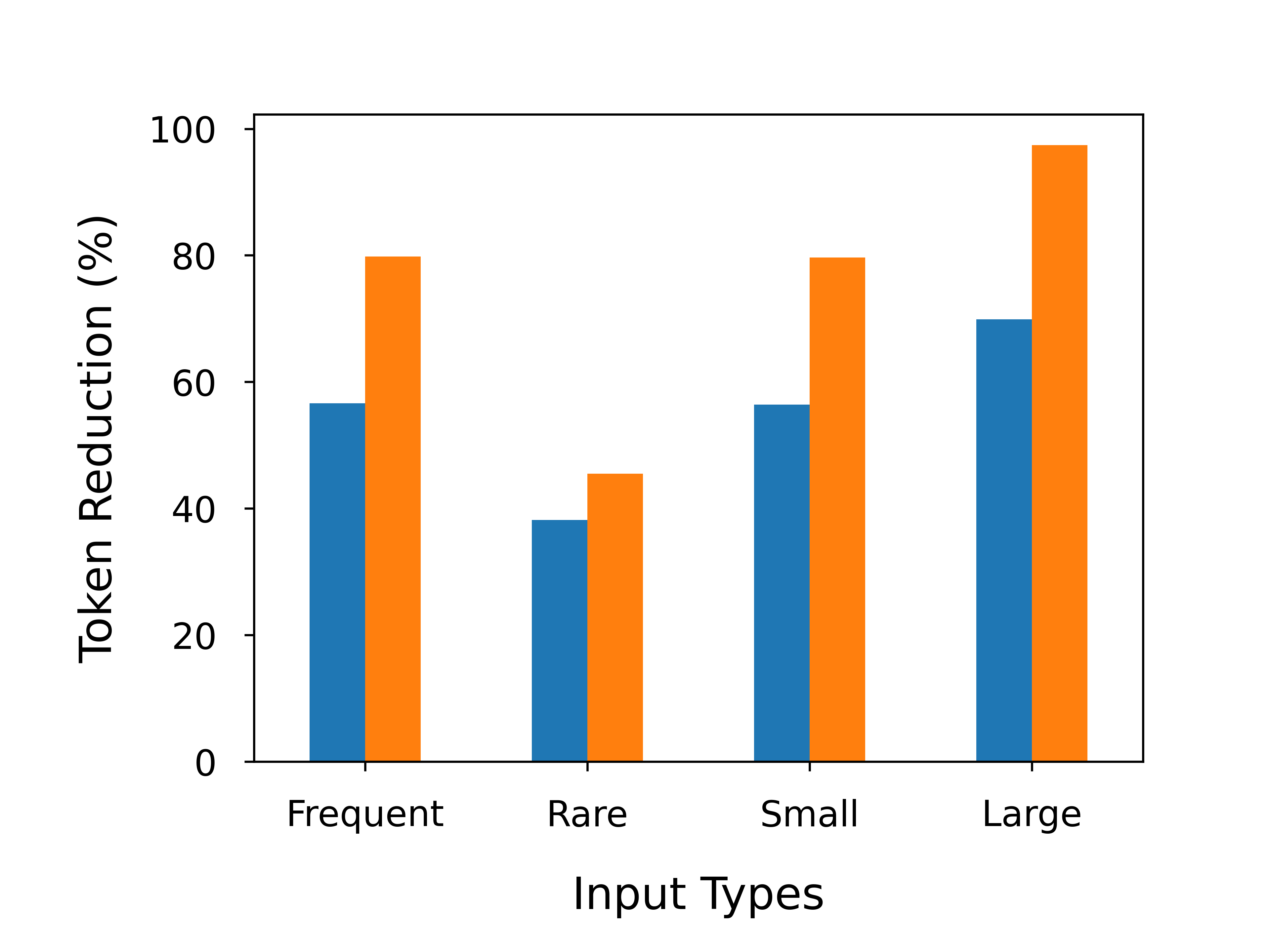}
  \caption{Token Reduction}
  \label{fig:reduction}
\end{subfigure}%
\begin{subfigure}{.33\textwidth}
  \centering
  \includegraphics[width=\linewidth]{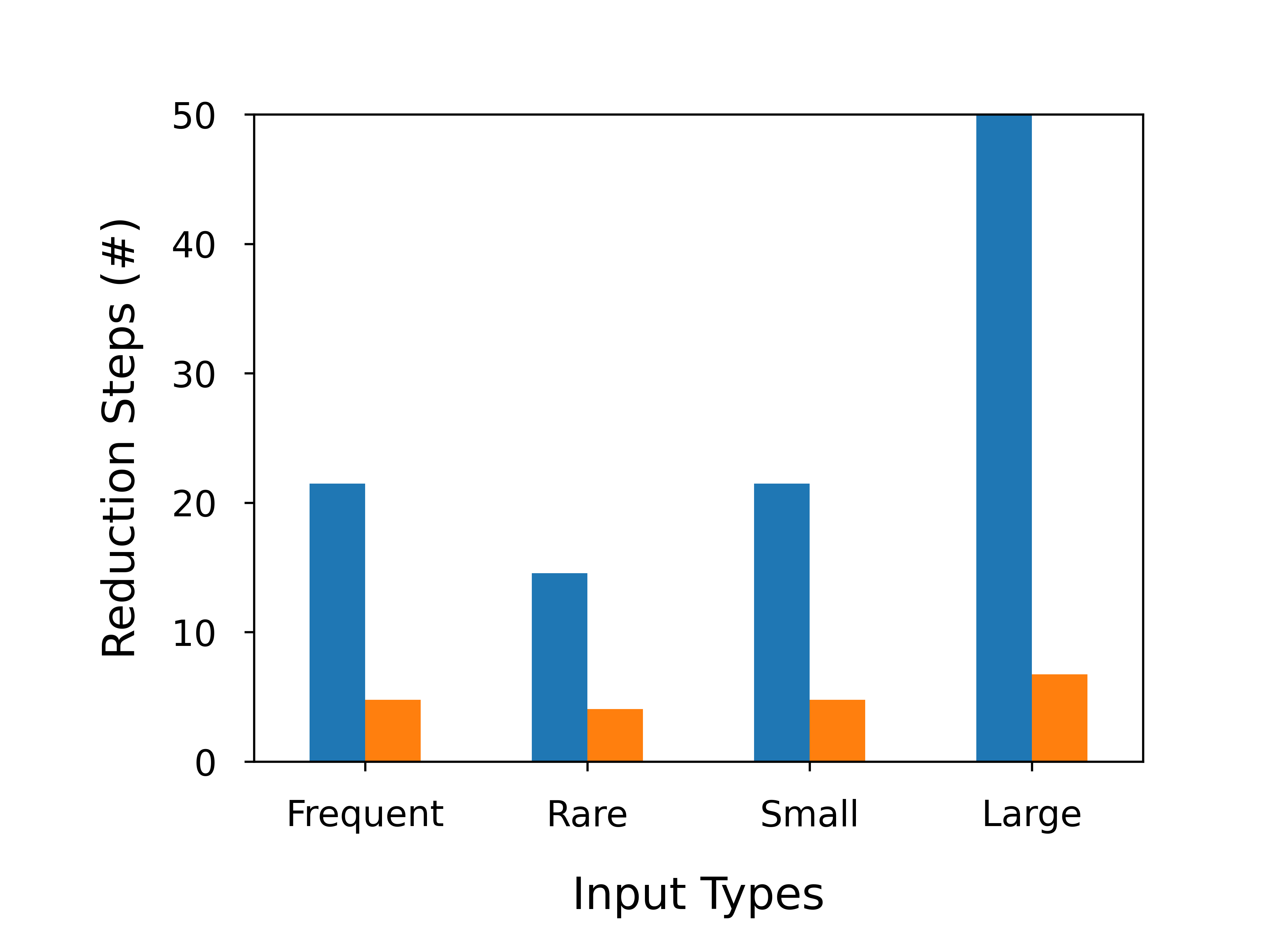}
  \caption{Reduction Steps}
  \label{fig:steps}
\end{subfigure}%
\begin{subfigure}{.33\textwidth}
  \centering
  \includegraphics[width=\linewidth]{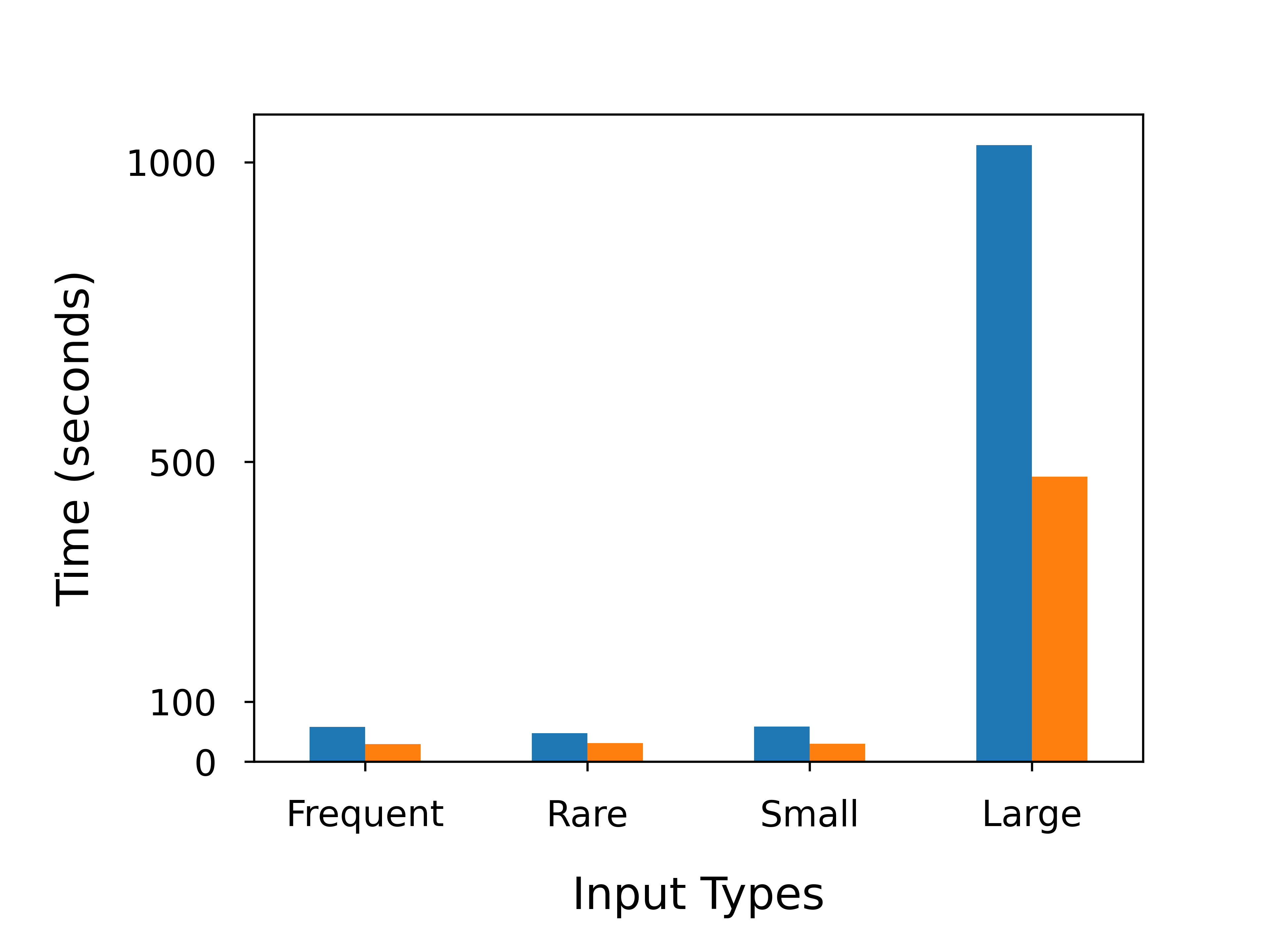}
  \caption{Reduction Time}
  \label{fig:time}
\end{subfigure}
\caption{Comparison between \sivanddd (\textcolor{blue}{\textbf{blue bar}}) and \sivandperses (\textcolor{orange}{\textbf{orange bar}}).}
\label{fig:comparison}
\end{figure*}

%% file: results/feature_summary.tex
\begin{table} 
    \begin{center}
        \def\arraystretch{1.2}
        \caption{Summary of input features in Top-10 methods.}
        \label{table:feature_summary}
        \resizebox{0.9\columnwidth}{!}{%
        \begin{tabular}{|c|c|c|c|c|c|}
            \hline
            \multirow{2}{*}{\textbf{Method}} & \multirow{2}{*}{\textbf{Model}} & \multicolumn{2}{c|}{\textbf{\sivanddd}} & \multicolumn{2}{c|}{\textbf{\sivandperses}} \\ \cline{3-6}
             & & \textbf{Candidate} & \textbf{Key} & \textbf{Candidate} & \textbf{Key} \\ \hline
            \hline
            
            \multirow{3}{*}{equals}
            & \ctv & 30  &  8  &  12  &  5 \\ \cline{2-6}
            & \cts & 28  &  8  &  10  &  5 \\ \cline{2-6}
            &(both)& 30  &  8  &  12  &  6 \\ \hline
            
            \multirow{3}{*}{main}
            & \ctv & 27  &  5  &  21  &  5 \\ \cline{2-6}
            & \cts & 27  &  5  &   4  &  3 \\ \cline{2-6}
            &(both)& 28  &  5  &  21  &  5 \\ \hline
            
            \multirow{3}{*}{setUp}
            & \ctv & 36  &  4  &  13  &  5 \\ \cline{2-6}
            & \cts & 27  &  4  &  13  &  1 \\ \cline{2-6}
            &(both)& 41  &  4  &  19  &  5 \\ \hline
            
            \multirow{3}{*}{onCreate}
            & \ctv & 31  &  5  &  20  &  4 \\ \cline{2-6}
            & \cts & 24  &  4  &  14  &  3 \\ \cline{2-6}
            &(both)& 34  &  5  &  24  &  4 \\ \hline
            
            \multirow{3}{*}{toString}
            & \ctv & 21  &  4  &  18  &  5 \\ \cline{2-6}
            & \cts & 22  &  3  &  18  &  2 \\ \cline{2-6}
            &(both)& 23  &  4  &  25  &  5 \\ \hline
            
            \multirow{3}{*}{run}
            & \ctv & 30  &  5  &  22  &  5 \\ \cline{2-6}
            & \cts & 31  &  6  &  13  &  2 \\ \cline{2-6}
            &(both)& 36  &  6  &  27  &  5 \\ \hline
            
            \multirow{3}{*}{hashCode}
            & \ctv & 13  &  5  &   6  &  5 \\ \cline{2-6}
            & \cts & 14  &  5  &  12  &  4 \\ \cline{2-6}
            &(both)& 15  &  5  &  13  &  5 \\ \hline
            
            \multirow{3}{*}{init}
            & \ctv & 30  &  3  &  29  &  3 \\ \cline{2-6}
            & \cts & 18  &  1  &  10  &  1 \\ \cline{2-6}
            &(both)& 35  &  3  &  33  &  3 \\ \hline
            
            \multirow{3}{*}{execute}
            & \ctv & 20  &  5  &  14  &  5 \\ \cline{2-6}
            & \cts & 16  &  3  &   7  &  3 \\ \cline{2-6}
            &(both)& 24  &  6  &  14  &  5 \\ \hline
            
            \multirow{3}{*}{get}
            & \ctv & 56  &  6  &  49  &  4 \\ \cline{2-6}
            & \cts & 23  &  0  &  16  &  0 \\ \cline{2-6}
            &(both)& 58  &  6  &  50  &  4 \\ \hline
            \hline
            \textbf{Top-10} 
            &(both)& \textbf{324} & \textbf{52} & \textbf{238} & \textbf{47} \\ \hline
        \end{tabular}%
        }
    \end{center}
\end{table}

%% file: results/feature_list.tex
\begin{table} 
    \begin{center}
        \def\arraystretch{1.2}
        \caption{Key input features in Top-10 methods.}
        \label{table:feature_list}
        \resizebox{0.98\columnwidth}{!}{%
        \begin{tabular}{|c|c|c|l| }
            \hline
            \textbf{Method} & \textbf{Model} & \textbf{Reduction} & \textbf{Key Input Features} \\ \hline
            \hline
            
            \multirow{4}{*}{equals}
            & \multirow{2}{*}{\ctv} 
            & \sivanddd & if, boolean, Object, o, obj, other, instanceof, Stock \\ \cline{3-4}
            & & \sivandperses & boolean, Object, return, o, obj \\ \cline{2-4}
            & \multirow{2}{*}{\cts} 
            & \sivanddd & if, boolean, Object, obj, other, o, instanceof, Stock \\ \cline{3-4}
            & & \sivandperses & boolean, Object, Override, o, obj \\ \hline
            
            \multirow{4}{*}{main}
            &  \multirow{2}{*}{\ctv} 
            & \sivanddd & args, void, String, Exception, throws \\ \cline{3-4}
            & & \sivandperses & void, String, args, System, Exception \\ \cline{2-4}
            &  \multirow{2}{*}{\cts} 
            & \sivanddd & args, void, String, Exception, throws \\ \cline{3-4}
            & & \sivandperses & void, String, args \\ \hline
            
            \multirow{4}{*}{setUp}
            &  \multirow{2}{*}{\ctv} 
            & \sivanddd & void, throws, Exception, setUp \\ \cline{3-4}
            & & \sivandperses & void, throws, Exception, super, setUp \\ \cline{2-4}
            &  \multirow{2}{*}{\cts} 
            & \sivanddd & void, throws, Exception, setUp \\ \cline{3-4}
            & & \sivandperses & void \\ \hline
            
            \multirow{4}{*}{onCreate}
            &  \multirow{2}{*}{\ctv} 
            & \sivanddd & void, savedInstanceState, Bundle, onCreate, if \\ \cline{3-4}
            & & \sivandperses & void, savedInstanceState, super, onCreate \\ \cline{2-4}
            &  \multirow{2}{*}{\cts} 
            & \sivanddd & void, savedInstanceState, onCreate, Bundle \\ \cline{3-4}
            & & \sivandperses & void, super, onCreate \\ \hline

            \multirow{4}{*}{toString}
            &  \multirow{2}{*}{\ctv} 
            & \sivanddd & String, if, toString, sb \\ \cline{3-4}
            & & \sivandperses & String, Override, StringBuilder, sb, return \\ \cline{2-4}
            &  \multirow{2}{*}{\cts} 
            & \sivanddd & String, toString, if \\ \cline{3-4}
            & & \sivandperses & String, return \\ \hline
            
            \multirow{4}{*}{run}
            &  \multirow{2}{*}{\ctv} 
            & \sivanddd & void, try, catch, 0, x \\ \cline{3-4}
            & & \sivandperses & void, Override, try, catch, x \\ \cline{2-4}
            &  \multirow{2}{*}{\cts} 
            & \sivanddd & void, try, catch, 0, Override, x \\ \cline{3-4}
            & & \sivandperses & void, Override \\ \hline
            
            \multirow{4}{*}{hashCode}
            &  \multirow{2}{*}{\ctv} 
            & \sivanddd & int, hashCode, 0, result, null \\ \cline{3-4}
            & & \sivandperses & int, Override, result, final, prime \\ \cline{2-4}
            &  \multirow{2}{*}{\cts} 
            & \sivanddd & int, hashCode, result, 0, null \\ \cline{3-4}
            & & \sivandperses & int, result, Override, prime \\ \hline
            
            \multirow{4}{*}{init}
            &  \multirow{2}{*}{\ctv} 
            & \sivanddd & void, throws, ServletException \\ \cline{3-4}
            & & \sivandperses & void, throws, ServletException \\ \cline{2-4}
            &  \multirow{2}{*}{\cts} 
            & \sivanddd & void \\ \cline{3-4}
            & & \sivandperses & void \\ \hline
            
            \multirow{4}{*}{execute}
            &  \multirow{2}{*}{\ctv} 
            & \sivanddd & void, throws, BuildException, execute, context \\ \cline{3-4}
            & & \sivandperses & void, throws, BuildException, super, execute \\ \cline{2-4}
            &  \multirow{2}{*}{\cts} 
            & \sivanddd & void, execute, super \\ \cline{3-4}
            & & \sivandperses & void, super, execute \\ \hline
            
            \multirow{4}{*}{get}
            &  \multirow{2}{*}{\ctv} 
            & \sivanddd & if, T, null, return, key, Object \\ \cline{3-4}
            & & \sivandperses & T, throw, key, return \\ \cline{2-4}
            &  \multirow{2}{*}{\cts} 
            & \sivanddd & None\\ \cline{3-4}
            & & \sivandperses & None \\ \hline
            
        \end{tabular}%
        }
    \end{center}
\end{table}

%% file: results/ex_explanation_main.tex

\begin{figure}
\textbf{(a) A sample input program:}
\vspace*{2mm}

\begin{lstlisting}
public static void f(String[] args) {
    System.setProperty(
        Constants.DUBBO_PROPERTIES_KEY, 
        "conf/dubbo.properties");
    Main.main(args);
}
\end{lstlisting}

\vspace*{2mm}
\textbf{(b) \sivand-Char reduced program:}
\vspace*{2mm}

\begin{lstlisting}
d f(Sg[]r){y(C,"");Main(ar);}
\end{lstlisting}

\vspace*{2mm}
\textbf{(c) \sivand-Token reduced program:}
\vspace*{2mm}

\begin{lstlisting}
void f(String[]args){("");(args);}
\end{lstlisting}

\vspace*{2mm}
\textbf{(d) \sivandperses reduced program:}
\vspace*{2mm}

\begin{lstlisting}
void f(String args) { }
\end{lstlisting}

\caption{A sample input program and corresponding reduced programs for different program reduction techniques.}
\label{code:explanation_main}
\end{figure}

%% file: results/key_adversarial.tex
\begin{table*} 
    \begin{center}
        \def\arraystretch{1.23}
        \caption{Adversarial evaluation with key input features.}
        \label{table:key_adversarial}
        \resizebox{0.76\textwidth}{!}{%
        \begin{tabular}{|c|c|c|c|c|c|c|}
            \hline
             \multirow{2}{*}{\textbf{Reduction}} & \multirow{2}{*}{\textbf{Model}} & \multirow{2}{*}{\textbf{Adversarial Set}} & \multirow{2}{*}{\textbf{\#Original}} & \multirow{2}{*}{\textbf{\#Transformed}} & \multicolumn{2}{c|}{\textbf{Misprediction}} \\ \cline{6-7}
             & & & & & \# & \% \\ \hline
            \hline
            
            \multirow{6}{*}{\sivanddd}
            & \multirow{3}{*}{\ctv} 
            & Original & \multirow{3}{*}{328} & 1148 & 135 & 11.76 \\ \cline{3-3}\cline{5-7}
            & & Key & & 722 & 107 & 14.82 \\ \cline{3-3}\cline{5-7}
            & & Reduced & & 379 & 117 & 30.87 \\ \cline{2-7}
            & \multirow{3}{*}{\cts} 
            & Original & \multirow{3}{*}{287} & 836 & 109 & 13.04 \\ \cline{3-3}\cline{5-7}
            & & Key & & 530 & 102 & 19.25 \\ \cline{3-3}\cline{5-7}
            & & Reduced & & 267 & 134 & 50.19 \\ \hline
            \hline
            
            \multirow{6}{*}{\sivandperses}
            & \multirow{3}{*}{\ctv} 
            & Original & \multirow{3}{*}{320} & 911 & 97 & 10.65 \\ \cline{3-3}\cline{5-7}
            & & Key & & 320 & 58 & 18.12 \\ \cline{3-3}\cline{5-7}
            & & Reduced & & 253 & 118 & 46.64 \\ \cline{2-7}
            & \multirow{3}{*}{\cts} 
            & Original & \multirow{3}{*}{280} & 658 & 93 & 14.13 \\ \cline{3-3}\cline{5-7}
            & & Key & & 211 & 80 & 37.91 \\ \cline{3-3}\cline{5-7}
            & & Reduced & & 160 & 104 & 65.00 \\ \hline
        \end{tabular}%
        }
    \end{center}
\end{table*}

%% file: validity.tex
\section{Threats to Validity and Future Plan}
\label{sec:results_threats}

\Part{Evaluation}. We evaluated our approach for \mnp task with two CI models, four input types of randomly selected input programs, and Top-$10$ most frequent method names. Despite our best effort, it is possible that experiments with different models, tasks, and datasets may produce different results. Our further plan includes a detailed study with a variety of models, tasks, and larger datasets. 

\Part{Challenges}. 
One challenge to integrating Perses is that it loads the model in each reduction step while \ddd loads the model once at the beginning of reduction. For a fair comparison between them, we only consider the program reduction time and ignore the model loading time. We are working on optimizing the model loading time for Perses.
Another challenge to integrating \ddd, when there are multiple subsets that hold the same target criteria, is that \ddd sometimes gets stuck at that point. To keep the reduction process working, we temporarily used a timer to stop the current step and jump to the next step.

%% file: conclusion.tex
\section{Conclusion}
\label{sec:conclusion}

In this paper, we apply the syntax-guided program reduction technique, \sivandperses, for reducing an input program while preserving the same prediction of the CI model.
The goal is to improve the transparency of models by extracting key input features from reduced programs.
We evaluate \sivandperses on two popular CI models across four types of input programs for the method name prediction task.
Our results suggest that the syntax-guided program reduction technique (\sivandperses) significantly outperforms the syntax-unaware program reduction technique (\sivanddd) in reducing different input programs.
Moreover, we extract key input features that CI models learn for a target label, by reducing some input programs of that label. The result shows that \sivandperses keeps more label-specific key input features in its syntax-guided reduced programs than in \sivanddd's syntax-unaware reduced programs.
We also observe that different program reduction techniques may provide additional explanations for a better understanding of a specific prediction.

%% file: impact.tex
\section{Data Availability Statement}
The proposed prediction-preserving program reduction framework for CI models and the corresponding reduced data using Perses and DD algorithms that support the findings of this study are openly available via Zenodo at \href{https://doi.org/10.5281/zenodo.6630188}{10.5281/zenodo.6630188} \cite{rabin2022artifact_perses}.

\section{Broader Impact}
In this section, we discuss the potential misuses and negative impacts of our approach. Our work focuses on improving the transparency of neural code intelligence models by reducing input programs with a syntax-guided technique and extracting key features that models learn from input programs. Therefore, in the negative scenario, attackers can try to force a model to make mispredictions by applying the key targeted adversarial attack (\cref{sec:results_adverserial}), thus may hamper the wider adoption of the model. However, our work also highlights multiple explanations for a prediction (\cref{sec:results_explanation}), which can motivate researchers to interpret and improve the robustness of a model against those malicious activities.

%% file: main.bbl

\begin{thebibliography}{28}


\ifx \showCODEN    \undefined \def \showCODEN     #1{\unskip}     \fi
\ifx \showDOI      \undefined \def \showDOI       #1{#1}\fi
\ifx \showISBNx    \undefined \def \showISBNx     #1{\unskip}     \fi
\ifx \showISBNxiii \undefined \def \showISBNxiii  #1{\unskip}     \fi
\ifx \showISSN     \undefined \def \showISSN      #1{\unskip}     \fi
\ifx \showLCCN     \undefined \def \showLCCN      #1{\unskip}     \fi
\ifx \shownote     \undefined \def \shownote      #1{#1}          \fi
\ifx \showarticletitle \undefined \def \showarticletitle #1{#1}   \fi
\ifx \showURL      \undefined \def \showURL       {\relax}        \fi
\providecommand\bibfield[2]{#2}
\providecommand\bibinfo[2]{#2}
\providecommand\natexlab[1]{#1}
\providecommand\showeprint[2][]{arXiv:#2}

\bibitem[Allamanis(2019)]%
        {allamanis2019duplication}
\bibfield{author}{\bibinfo{person}{Miltiadis Allamanis}.}
  \bibinfo{year}{2019}\natexlab{}.
\newblock \showarticletitle{The Adverse Effects of Code Duplication in Machine
  Learning Models of Code}. In \bibinfo{booktitle}{\emph{Proceedings of the ACM
  SIGPLAN International Symposium on New Ideas, New Paradigms, and Reflections
  on Programming and Software}} (Athens, Greece)
  \emph{(\bibinfo{series}{Onward! 2019})}. \bibinfo{publisher}{ACM New York,
  NY, USA}, \bibinfo{pages}{143--153}.
\newblock
\urldef\tempurl%
\url{https://doi.org/10.1145/3359591.3359735}
\showDOI{\tempurl}


\bibitem[Allamanis et~al\mbox{.}(2014)]%
        {allamanis2014learning}
\bibfield{author}{\bibinfo{person}{Miltiadis Allamanis},
  \bibinfo{person}{Earl~T. Barr}, \bibinfo{person}{Christian Bird}, {and}
  \bibinfo{person}{Charles Sutton}.} \bibinfo{year}{2014}\natexlab{}.
\newblock \showarticletitle{Learning Natural Coding Conventions}. In
  \bibinfo{booktitle}{\emph{Proceedings of the 22nd ACM SIGSOFT International
  Symposium on Foundations of Software Engineering}} (Hong Kong, China)
  \emph{(\bibinfo{series}{FSE 2014})}. \bibinfo{publisher}{Association for
  Computing Machinery}, \bibinfo{address}{New York, NY, USA},
  \bibinfo{pages}{281--293}.
\newblock
\urldef\tempurl%
\url{https://doi.org/10.1145/2635868.2635883}
\showDOI{\tempurl}


\bibitem[Allamanis et~al\mbox{.}(2015)]%
        {allamanis2015suggesting}
\bibfield{author}{\bibinfo{person}{Miltiadis Allamanis},
  \bibinfo{person}{Earl~T. Barr}, \bibinfo{person}{Christian Bird}, {and}
  \bibinfo{person}{Charles Sutton}.} \bibinfo{year}{2015}\natexlab{}.
\newblock \showarticletitle{Suggesting Accurate Method and Class Names}. In
  \bibinfo{booktitle}{\emph{Proceedings of the 10th Joint Meeting on
  Foundations of Software Engineering}} (Bergamo, Italy)
  \emph{(\bibinfo{series}{ESEC/FSE 2015})}. \bibinfo{publisher}{Association for
  Computing Machinery}, \bibinfo{address}{New York, NY, USA},
  \bibinfo{pages}{38--49}.
\newblock
\urldef\tempurl%
\url{https://doi.org/10.1145/2786805.2786849}
\showDOI{\tempurl}


\bibitem[Allamanis et~al\mbox{.}(2018a)]%
        {allamanis2018survey}
\bibfield{author}{\bibinfo{person}{Miltiadis Allamanis},
  \bibinfo{person}{Earl~T. Barr}, \bibinfo{person}{Premkumar Devanbu}, {and}
  \bibinfo{person}{Charles Sutton}.} \bibinfo{year}{2018}\natexlab{a}.
\newblock \showarticletitle{A Survey of Machine Learning for Big Code and
  Naturalness}. In \bibinfo{booktitle}{\emph{{{A}{C}{M}} Computing Surveys}},
  Vol.~\bibinfo{volume}{51}. \bibinfo{publisher}{Association for Computing
  Machinery}, \bibinfo{address}{New York, NY, USA}, Article
  \bibinfo{articleno}{81}, \bibinfo{numpages}{37}~pages.
\newblock
\urldef\tempurl%
\url{https://doi.org/10.1145/3212695}
\showDOI{\tempurl}


\bibitem[Allamanis et~al\mbox{.}(2018b)]%
        {allamanis2018ggnn}
\bibfield{author}{\bibinfo{person}{Miltiadis Allamanis}, \bibinfo{person}{Marc
  Brockschmidt}, {and} \bibinfo{person}{Mahmoud Khademi}.}
  \bibinfo{year}{2018}\natexlab{b}.
\newblock \showarticletitle{Learning to Represent Programs with Graphs}. In
  \bibinfo{booktitle}{\emph{International Conference on Learning
  Representations}} (Vancouver, BC, Canada) \emph{(\bibinfo{series}{ICLR
  2018})}. \bibinfo{publisher}{OpenReview.net}, \bibinfo{address}{Open Access}.
\newblock
\urldef\tempurl%
\url{https://openreview.net/forum?id=BJOFETxR-}
\showURL{%
\tempurl}


\bibitem[Allamanis et~al\mbox{.}(2016)]%
        {allamanis2016summarization}
\bibfield{author}{\bibinfo{person}{Miltiadis Allamanis}, \bibinfo{person}{Hao
  Peng}, {and} \bibinfo{person}{Charles~A. Sutton}.}
  \bibinfo{year}{2016}\natexlab{}.
\newblock \showarticletitle{A Convolutional Attention Network for Extreme
  Summarization of Source Code}. In \bibinfo{booktitle}{\emph{Proceedings of
  the 33nd International Conference on Machine Learning}} (New York, NY, USA)
  \emph{(\bibinfo{series}{ICML 2016})}. \bibinfo{publisher}{Proceedings of
  Machine Learning Research (PMLR)}, \bibinfo{address}{Open Access},
  \bibinfo{pages}{2091--2100}.
\newblock
\urldef\tempurl%
\url{http://proceedings.mlr.press/v48/allamanis16.html}
\showURL{%
\tempurl}


\bibitem[Alon et~al\mbox{.}(2019a)]%
        {alon2019code2seq}
\bibfield{author}{\bibinfo{person}{Uri Alon}, \bibinfo{person}{Omer Levy},
  {and} \bibinfo{person}{Eran Yahav}.} \bibinfo{year}{2019}\natexlab{a}.
\newblock \showarticletitle{code2seq: Generating Sequences from Structured
  Representations of Code}. In \bibinfo{booktitle}{\emph{International
  Conference on Learning Representations}} (New Orleans, Louisiana, United
  States) \emph{(\bibinfo{series}{ICLR 2019})}.
  \bibinfo{publisher}{OpenReview.net}, \bibinfo{address}{Open Access}.
\newblock
\urldef\tempurl%
\url{https://openreview.net/forum?id=H1gKYo09tX}
\showURL{%
\tempurl}


\bibitem[Alon et~al\mbox{.}(2019b)]%
        {alon2019code2vec}
\bibfield{author}{\bibinfo{person}{Uri Alon}, \bibinfo{person}{Meital
  Zilberstein}, \bibinfo{person}{Omer Levy}, {and} \bibinfo{person}{Eran
  Yahav}.} \bibinfo{year}{2019}\natexlab{b}.
\newblock \showarticletitle{Code2vec: Learning Distributed Representations of
  Code}. In \bibinfo{booktitle}{\emph{Proceedings of the ACM on Programming
  Languages}} (Cascais, Portugal) \emph{(\bibinfo{series}{PACMPL 2019},
  Vol.~\bibinfo{volume}{3})}. \bibinfo{publisher}{Association for Computing
  Machinery}, \bibinfo{address}{New York, NY, USA},
  \bibinfo{pages}{40:1--40:29}.
\newblock
\urldef\tempurl%
\url{https://doi.org/10.1145/3290353}
\showDOI{\tempurl}


\bibitem[Bui et~al\mbox{.}(2019)]%
        {bui2019autofocus}
\bibfield{author}{\bibinfo{person}{Nghi D.~Q. Bui}, \bibinfo{person}{Yijun Yu},
  {and} \bibinfo{person}{Lingxiao Jiang}.} \bibinfo{year}{2019}\natexlab{}.
\newblock \showarticletitle{AutoFocus: Interpreting Attention-Based Neural
  Networks by Code Perturbation}. In \bibinfo{booktitle}{\emph{Proceedings of
  the 34th IEEE/ACM International Conference on Automated Software
  Engineering}} (San Diego, CA, USA) \emph{(\bibinfo{series}{ASE 2019})}.
  \bibinfo{publisher}{IEEE Press}, \bibinfo{address}{New York, NY, USA},
  \bibinfo{pages}{38--41}.
\newblock
\urldef\tempurl%
\url{https://doi.org/10.1109/ASE.2019.00014}
\showDOI{\tempurl}


\bibitem[Chen and Monperrus(2019)]%
        {chen2019embeddings}
\bibfield{author}{\bibinfo{person}{Zimin Chen} {and} \bibinfo{person}{Martin
  Monperrus}.} \bibinfo{year}{2019}\natexlab{}.
\newblock \bibinfo{title}{A Literature Study of Embeddings on Source Code}.
\newblock
\newblock
\showeprint[arxiv]{1904.03061}~[cs.LG]
\urldef\tempurl%
\url{https://arxiv.org/abs/1904.03061}
\showURL{%
\tempurl}


\bibitem[Compton et~al\mbox{.}(2020)]%
        {compton2020obfuscation}
\bibfield{author}{\bibinfo{person}{Rhys Compton}, \bibinfo{person}{Eibe Frank},
  \bibinfo{person}{Panos Patros}, {and} \bibinfo{person}{Abigail Koay}.}
  \bibinfo{year}{2020}\natexlab{}.
\newblock \showarticletitle{Embedding Java Classes with Code2vec: Improvements
  from Variable Obfuscation}. In \bibinfo{booktitle}{\emph{Proceedings of the
  17th International Conference on Mining Software Repositories}} (Seoul,
  Republic of Korea) \emph{(\bibinfo{series}{MSR 2020})}.
  \bibinfo{publisher}{Association for Computing Machinery},
  \bibinfo{address}{New York, NY, USA}, \bibinfo{pages}{243--253}.
\newblock
\urldef\tempurl%
\url{https://doi.org/10.1145/3379597.3387445}
\showDOI{\tempurl}


\bibitem[Hellendoorn et~al\mbox{.}(2018)]%
        {hellendoorn2018type}
\bibfield{author}{\bibinfo{person}{Vincent~J. Hellendoorn},
  \bibinfo{person}{Christian Bird}, \bibinfo{person}{Earl~T. Barr}, {and}
  \bibinfo{person}{Miltiadis Allamanis}.} \bibinfo{year}{2018}\natexlab{}.
\newblock \showarticletitle{Deep Learning Type Inference}. In
  \bibinfo{booktitle}{\emph{Proceedings of the 26th ACM Joint Meeting on
  European Software Engineering Conference and Symposium on the Foundations of
  Software Engineering}} (Lake Buena Vista, FL, USA)
  \emph{(\bibinfo{series}{ESEC/FSE 2018})}. \bibinfo{publisher}{Association for
  Computing Machinery}, \bibinfo{address}{New York, NY, USA},
  \bibinfo{pages}{152--162}.
\newblock
\urldef\tempurl%
\url{https://doi.org/10.1145/3236024.3236051}
\showDOI{\tempurl}


\bibitem[Kang et~al\mbox{.}(2019)]%
        {kang2019generalizability}
\bibfield{author}{\bibinfo{person}{Hong~Jin Kang},
  \bibinfo{person}{Tegawend\'{e}~F. Bissyand\'{e}}, {and}
  \bibinfo{person}{David Lo}.} \bibinfo{year}{2019}\natexlab{}.
\newblock \showarticletitle{Assessing the Generalizability of Code2vec Token
  Embeddings}. In \bibinfo{booktitle}{\emph{Proceedings of the 34th IEEE/ACM
  International Conference on Automated Software Engineering}} (San Diego, CA,
  USA) \emph{(\bibinfo{series}{ASE 2019})}. \bibinfo{publisher}{IEEE Press},
  \bibinfo{address}{New York, NY, USA}, \bibinfo{pages}{1--12}.
\newblock
\urldef\tempurl%
\url{https://doi.org/10.1109/ASE.2019.00011}
\showDOI{\tempurl}


\bibitem[Rabin and Alipour(2020)]%
        {rabin2020evaluation}
\bibfield{author}{\bibinfo{person}{Md~Rafiqul~Islam Rabin} {and}
  \bibinfo{person}{Mohammad~Amin Alipour}.} \bibinfo{year}{2020}\natexlab{}.
\newblock \bibinfo{title}{Evaluation of Generalizability of Neural Program
  Analyzers under Semantic-Preserving Transformations}.
\newblock
\newblock
\showeprint[arxiv]{2004.07313}~[cs.SE]
\urldef\tempurl%
\url{https://arxiv.org/abs/2004.07313}
\showURL{%
\tempurl}


\bibitem[Rabin and Alipour(2021)]%
        {rabin2021code2snapshot}
\bibfield{author}{\bibinfo{person}{Md~Rafiqul~Islam Rabin} {and}
  \bibinfo{person}{Mohammad~Amin Alipour}.} \bibinfo{year}{2021}\natexlab{}.
\newblock \bibinfo{title}{Code2Snapshot: Using Code Snapshots for Learning
  Representations of Source Code}.
\newblock
\newblock
\showeprint[arxiv]{2111.01097}~[cs.SE]
\urldef\tempurl%
\url{https://arxiv.org/abs/2111.01097}
\showURL{%
\tempurl}


\bibitem[Rabin et~al\mbox{.}(2021a)]%
        {rabin2021generalizability}
\bibfield{author}{\bibinfo{person}{Md~Rafiqul~Islam Rabin},
  \bibinfo{person}{Nghi~D.Q. Bui}, \bibinfo{person}{Ke Wang},
  \bibinfo{person}{Yijun Yu}, \bibinfo{person}{Lingxiao Jiang}, {and}
  \bibinfo{person}{Mohammad~Amin Alipour}.} \bibinfo{year}{2021}\natexlab{a}.
\newblock \showarticletitle{On the generalizability of Neural Program Models
  with respect to semantic-preserving program transformations}. In
  \bibinfo{booktitle}{\emph{Information and Software Technology}},
  Vol.~\bibinfo{volume}{135}. \bibinfo{publisher}{Elsevier},
  \bibinfo{address}{Amsterdam, Netherlands}, \bibinfo{pages}{106552}.
\newblock
\urldef\tempurl%
\url{https://doi.org/10.1016/j.infsof.2021.106552}
\showDOI{\tempurl}


\bibitem[Rabin et~al\mbox{.}(2021b)]%
        {rabin2021artifact_dd}
\bibfield{author}{\bibinfo{person}{Md~Rafiqul~Islam Rabin},
  \bibinfo{person}{Vincent~J. Hellendoorn}, {and}
  \bibinfo{person}{Mohammad~Amin Alipour}.} \bibinfo{year}{2021}\natexlab{b}.
\newblock \showarticletitle{Artifact for Article (SIVAND): Understanding Neural
  Code Intelligence Through Program Simplification}.
\newblock \bibinfo{journal}{\emph{{ACM} Digital Library, ESEC/FSE}}
  (\bibinfo{year}{2021}).
\newblock
\urldef\tempurl%
\url{https://doi.org/10.1145/3462296}
\showURL{%
\tempurl}


\bibitem[Rabin et~al\mbox{.}(2021c)]%
        {rabin2021sivand}
\bibfield{author}{\bibinfo{person}{Md~Rafiqul~Islam Rabin},
  \bibinfo{person}{Vincent~J. Hellendoorn}, {and}
  \bibinfo{person}{Mohammad~Amin Alipour}.} \bibinfo{year}{2021}\natexlab{c}.
\newblock \showarticletitle{Understanding Neural Code Intelligence through
  Program Simplification}. In \bibinfo{booktitle}{\emph{Proceedings of the 29th
  ACM Joint Meeting on European Software Engineering Conference and Symposium
  on the Foundations of Software Engineering}} (Athens, Greece)
  \emph{(\bibinfo{series}{ESEC/FSE 2021})}. \bibinfo{publisher}{Association for
  Computing Machinery}, \bibinfo{address}{New York, NY, USA},
  \bibinfo{pages}{441--452}.
\newblock
\urldef\tempurl%
\url{https://doi.org/10.1145/3468264.3468539}
\showDOI{\tempurl}


\bibitem[Rabin et~al\mbox{.}(2022)]%
        {rabin2022artifact_perses}
\bibfield{author}{\bibinfo{person}{Md~Rafiqul~Islam Rabin},
  \bibinfo{person}{Aftab Hussain}, {and} \bibinfo{person}{Mohammad~Amin
  Alipour}.} \bibinfo{year}{2022}\natexlab{}.
\newblock \showarticletitle{Artifact for Article (CI-DD-Perses): Syntax-Guided
  Program Reduction for Understanding Neural Code Intelligence Models}.
\newblock \bibinfo{journal}{\emph{Zenodo, MAPS}} (\bibinfo{year}{2022}).
\newblock
\urldef\tempurl%
\url{https://doi.org/10.5281/zenodo.6630188}
\showURL{%
\tempurl}


\bibitem[Rabin et~al\mbox{.}(2021d)]%
        {rabin2021memorization}
\bibfield{author}{\bibinfo{person}{Md~Rafiqul~Islam Rabin},
  \bibinfo{person}{Aftab Hussain}, \bibinfo{person}{Mohammad~Amin Alipour},
  {and} \bibinfo{person}{Vincent~J. Hellendoorn}.}
  \bibinfo{year}{2021}\natexlab{d}.
\newblock \bibinfo{title}{Memorization and Generalization in Neural Code
  Intelligence Models}.
\newblock
\newblock
\showeprint[arxiv]{2106.08704}~[cs.LG]
\urldef\tempurl%
\url{https://arxiv.org/abs/2106.08704}
\showURL{%
\tempurl}


\bibitem[Rabin et~al\mbox{.}(2020)]%
        {rabin2020demystifying}
\bibfield{author}{\bibinfo{person}{Md~Rafiqul~Islam Rabin},
  \bibinfo{person}{Arjun Mukherjee}, \bibinfo{person}{Omprakash Gnawali}, {and}
  \bibinfo{person}{Mohammad~Amin Alipour}.} \bibinfo{year}{2020}\natexlab{}.
\newblock \showarticletitle{Towards Demystifying Dimensions of Source Code
  Embeddings}. In \bibinfo{booktitle}{\emph{Proceedings of the 1st ACM SIGSOFT
  International Workshop on Representation Learning for Software Engineering
  and Program Languages}} (Virtual, USA) \emph{(\bibinfo{series}{RL+SE\&PL
  2020})}. \bibinfo{publisher}{Association for Computing Machinery},
  \bibinfo{address}{New York, NY, USA}, \bibinfo{pages}{29--38}.
\newblock
\urldef\tempurl%
\url{https://doi.org/10.1145/3416506.3423580}
\showDOI{\tempurl}


\bibitem[Rabin et~al\mbox{.}(2019)]%
        {rabin2019tnpa}
\bibfield{author}{\bibinfo{person}{Md~Rafiqul~Islam Rabin}, \bibinfo{person}{Ke
  Wang}, {and} \bibinfo{person}{Mohammad~Amin Alipour}.}
  \bibinfo{year}{2019}\natexlab{}.
\newblock \showarticletitle{Testing Neural Program Analyzers}.
\newblock \bibinfo{journal}{\emph{34th IEEE/ACM International Conference on
  Automated Software Engineering (Late Breaking Results-Track)}}
  (\bibinfo{year}{2019}).
\newblock
\urldef\tempurl%
\url{https://arxiv.org/abs/1908.10711}
\showURL{%
\tempurl}


\bibitem[Sharma et~al\mbox{.}(2021)]%
        {sharma2021survey}
\bibfield{author}{\bibinfo{person}{Tushar Sharma}, \bibinfo{person}{Maria
  Kechagia}, \bibinfo{person}{Stefanos Georgiou}, \bibinfo{person}{Rohit
  Tiwari}, {and} \bibinfo{person}{Federica Sarro}.}
  \bibinfo{year}{2021}\natexlab{}.
\newblock \bibinfo{title}{A Survey on Machine Learning Techniques for Source
  Code Analysis}.
\newblock
\newblock
\showeprint[arxiv]{2110.09610}~[cs.SE]
\urldef\tempurl%
\url{https://arxiv.org/abs/2110.09610}
\showURL{%
\tempurl}


\bibitem[Sun et~al\mbox{.}(2018)]%
        {sun2018perses}
\bibfield{author}{\bibinfo{person}{Chengnian Sun}, \bibinfo{person}{Yuanbo Li},
  \bibinfo{person}{Qirun Zhang}, \bibinfo{person}{Tianxiao Gu}, {and}
  \bibinfo{person}{Zhendong Su}.} \bibinfo{year}{2018}\natexlab{}.
\newblock \showarticletitle{Perses: Syntax-Guided Program Reduction}. In
  \bibinfo{booktitle}{\emph{Proceedings of the 40th International Conference on
  Software Engineering}} (Gothenburg, Sweden) \emph{(\bibinfo{series}{ICSE
  2018})}. \bibinfo{publisher}{Association for Computing Machinery},
  \bibinfo{address}{New York, NY, USA}, \bibinfo{pages}{361--371}.
\newblock
\urldef\tempurl%
\url{https://doi.org/10.1145/3180155.3180236}
\showDOI{\tempurl}


\bibitem[Suneja et~al\mbox{.}(2021)]%
        {suneja2021probing}
\bibfield{author}{\bibinfo{person}{Sahil Suneja}, \bibinfo{person}{Yunhui
  Zheng}, \bibinfo{person}{Yufan Zhuang}, \bibinfo{person}{Jim~A. Laredo},
  {and} \bibinfo{person}{Alessandro Morari}.} \bibinfo{year}{2021}\natexlab{}.
\newblock \showarticletitle{Probing Model Signal-Awareness via
  Prediction-Preserving Input Minimization}. In
  \bibinfo{booktitle}{\emph{Proceedings of the 29th ACM Joint Meeting on
  European Software Engineering Conference and Symposium on the Foundations of
  Software Engineering}} (Athens, Greece) \emph{(\bibinfo{series}{ESEC/FSE
  2021})}. \bibinfo{publisher}{Association for Computing Machinery},
  \bibinfo{address}{New York, NY, USA}, \bibinfo{pages}{945–955}.
\newblock
\urldef\tempurl%
\url{https://doi.org/10.1145/3468264.3468545}
\showDOI{\tempurl}


\bibitem[Wang et~al\mbox{.}(2021)]%
        {wang2021demystifying}
\bibfield{author}{\bibinfo{person}{Yu Wang}, \bibinfo{person}{Fengjuan Gao},
  {and} \bibinfo{person}{Linzhang Wang}.} \bibinfo{year}{2021}\natexlab{}.
\newblock \bibinfo{title}{Demystifying code summarization models}.
\newblock
\newblock
\showeprint[arxiv]{2102.04625}~[cs.LG]
\urldef\tempurl%
\url{https://arxiv.org/abs/2102.04625}
\showURL{%
\tempurl}


\bibitem[Yefet et~al\mbox{.}(2020)]%
        {yefet2020adversarial}
\bibfield{author}{\bibinfo{person}{Noam Yefet}, \bibinfo{person}{Uri Alon},
  {and} \bibinfo{person}{Eran Yahav}.} \bibinfo{year}{2020}\natexlab{}.
\newblock \showarticletitle{Adversarial Examples for Models of Code}. In
  \bibinfo{booktitle}{\emph{Proceedings of the ACM on Programming Languages}}
  (Virtual, USA) \emph{(\bibinfo{series}{PACMPL 2020},
  Vol.~\bibinfo{volume}{4})}. \bibinfo{publisher}{Association for Computing
  Machinery}, \bibinfo{address}{New York, NY, USA},
  \bibinfo{pages}{162:1--162:30}.
\newblock
\urldef\tempurl%
\url{https://doi.org/10.1145/3428230}
\showDOI{\tempurl}


\bibitem[Zeller and Hildebrandt(2002)]%
        {zeller2002dd}
\bibfield{author}{\bibinfo{person}{Andreas Zeller} {and} \bibinfo{person}{Ralf
  Hildebrandt}.} \bibinfo{year}{2002}\natexlab{}.
\newblock \showarticletitle{Simplifying and Isolating Failure-Inducing Input}.
  In \bibinfo{booktitle}{\emph{IEEE Transactions on Software Engineering}}
  (Virtual), Vol.~\bibinfo{volume}{28}. \bibinfo{publisher}{IEEE Press},
  \bibinfo{address}{New York, NY, USA}, \bibinfo{pages}{183--200}.
\newblock
\urldef\tempurl%
\url{https://doi.org/10.1109/32.988498}
\showDOI{\tempurl}


\end{thebibliography}
